\documentclass[twocolumn,showpacs,prd,floatfix]{revtex4}
\usepackage{psfig} \usepackage{amssymb} \usepackage{amsmath}

\begin{document}
\newcommand{\etal}{{et al.}~}
\newcommand{\be}{\begin{equation}}
\newcommand{\ee}{\end{equation}}
\newcommand{\ba}{\begin{eqnarray}}
\newcommand{\ea}{\end{eqnarray}}
\newcommand{\bd}{\mbox{\bf d}}
\newcommand{\bv}{\mbox{\bf v}}
\newcommand{\bg}{\mbox{\bf g}}
\newcommand{\lb}{{\left<\right.}}
\newcommand{\rb}{{\left.\right>}}
\newcommand{\kms}{\ensuremath{\hbox{km} \hbox{sec}^{-1}}}
\newcommand{\hmpc}{\ensuremath{h^{-1}\,\hbox{Mpc}}}
\def\omegam{{\Omega_{\rm m}}}
\def\msun{\rm {M_{\odot}}}

\input{psfig.sty}

\title[The LG velocity] {Is the misalignment of the Local Group velocity and the 2MRS dipole typical in a $\Lambda$CDM \& halo model?}
\author{Pirin Erdo{\u g}du$^{1,2}$\thanks{pirin@star.ucl.ac.uk} and Ofer Lahav$^{1}$\thanks{lahav@star.ucl.ac.uk}}

\affiliation{$^{1}$Department of Physics and Astronomy, University College London, London, WC1E 6BT}
\affiliation{$^{2}$Department of Science and Engineering, American University of Kuwait, P.O. Box 3323, Safat 13034, Kuwait} 

\date{\today}

\begin{abstract}
We predict the acceleration of the Local Group generated by the 2MASS
Redshift Survey (2MRS) within the framework of $\Lambda$CDM and the
halo model of galaxies.  We show that as the galaxy fluctuations
derived from the halo model have more power on small scales compared
with the mass fluctuations, the misalignment angle between the CMB
velocity vector and the 2MRS dipole is in reasonable agreement
with the observed 21$^\circ$.  This statistical analysis suggests that
it is not necessary to invoke a hypothetical nearby galaxy or a distant
cluster to explain this misalignment.
\end{abstract}

\pacs{98.56}

\maketitle

\section{Introduction}

In the late 1980s and early 1990s, the near alignment of the $IRAS$
and the optical dipoles with the Local Group velocity relative to the cosmic microwave background (CMB, \cite{Yahil, Meiksin, Harmon, Villumsen, lahav1, lynden, Strauss, Webster})
were used to verify the gravitational instability picture of structure
formation and the scale of convergence was used to assess the validity
of the then popular Standard Cold Dark Matter (SCDM) model 
(\cite{LKH} hereafter (LKH) and \cite{JVW}, hereafter JVW).

The distance 
at which most of the Local Group (LG) peculiar velocity is generated 
({\it the convergence depth}) has remained a 
contentious issue for the past twenty five years.
The observed misalignment between the vectors derived from the galaxy survey and the CMB has been attributed to several factors:
\begin{enumerate}
  \setlength{\itemsep}{1pt}
  \setlength{\parskip}{0pt}
  \setlength{\parsep}{0pt}
\item The linear perturbation theory of density fluctuations is correct
only to first order density contrast, $\delta$.  There may be contributions to the LG
dipole from small scales which would cause gravity and the velocity
vectors to misalign. 
\item The selection effects of the
surveys from which these vectors are calculated will increase the shot noise-error especially at large
distances causing misalignments.
\item There may be uncertainties in the assumptions in galaxy
formation and clustering. For example, the mass-to-light ratios might
differ according to type and/or vary with luminosity or the galaxy/cluster
biasing might be non-linear and/or scale dependent.
\item There may be a significant contribution to the LG dipole from
structure further away than the maximum distance of the surveys studied.
\item The direction of the LG dipole may be affected by nearby
galaxies or distant clusters behind the {\it Zone of Avoidance} (ZoA), which are
not sampled by most surveys.
\end{enumerate}
The estimates of the convergence depth are highly sensitive to the type of data and the analysis performed. For example, the convergence depth estimated from galaxy clusters consistently disagree with the estimates from galaxy surveys (e.g. see \cite{erdogdu1} and the references therein).  Moreover, the treatment of the ZoA plays an important factor in dipole determinations. Exclusion of bright galaxies changes the dipole misalignment direction considerably (from 21$^\circ$ to 14$^\circ$, \cite{erdogdu1}). These bright galaxies lie within the ZoA and are excluded from most dipole analyses. This also highlights the vital role non-linear dynamics introduced by nearby objects play in dipole considerations. The analysis of the
convergence of the dipole is further complicated by the redshift distortions on small and large scales which introduce systematic
errors \cite{Kaiser}.

Recently, the new dipole determinations from the 2MASS Redshift Survey (hereafter 2MRS, \cite{erdogdu1, erdogdu2, lavaux, crook}
brought up once again the question of alignment and convergence.
For example, \cite{Loeb} attempted to fix the observed misalignment by
placing a hypothetical object behind the ZoA and \cite{Chodorowski} tried to 
mitigate the non-linear and shot noise effects by removing nearby objects in order to increase the correlation between the dipole determined from the CMB and the 2MRS.
Furthermore, two recent studies of the peculiar velocity surveys indicated a large convergence depth. 
\cite{Kashlinsky} found a strong bulk flow on scales 
out to 300 $h^{-1} {\rm Mpc}$ using peculiar velocity data from an X-ray cluster sample. \cite{Watkins} reported a large-scale bulk flow beyond 50 $h^{-1} {\rm Mpc}$ from a comprehensive compilation of peculiar velocity data. Both these papers claimed that these convergence depths, determined by 
their studies, were difficult to explain within the framework of standard $\Lambda$CDM model of cosmology.

Here we take a different approach, updating the LKH and JVW
calculations to include two major developments in cosmology: the
$\Lambda$CDM concordance model (replacing the old SCDM and a halo
model for the galaxy power spectrum (replacing the naive linear
biasing model). In the next section, we consider the expected
misalignment between the peculiar velocity of the LG and the
acceleration vector estimates.  We expect that the $\Lambda$CDM model
will give a poorer misalignment compared to the SCDM model since it
has a larger coherence length. Adding the halo model for galaxy
biasing will further increase the power of density fluctuations on
small scales and therefore the misalignment angle.  In Section 3, we
consider the convergence of the acceleration from the 2MRS survey.  In
Section 4, we show that the observed convergence is highly probable in
our model. In the last section we discuss limitations of our
calculations and future work.

\section{Alignment}

The most popular mechanism for the formation of large-scale structure
and motions in the Universe is the gravitational growth of primordial
density perturbations.  According to this paradigm the peculiar
acceleration vector ${\bf g}({\bf r})$ induced by the matter
distribution around position ${\bf r}$ is related to the mass by
\begin{equation}
{\bf g}({\bf r})= G\bar{\rho} \int\limits_{\bf r}^{\infty} d^3{\bf
r}^{\prime} \delta_m({{\bf r}^{\prime}}) \frac{{\bf r}^{\prime}-{\bf
r}}{|{\bf r}^{\prime}-{\bf r}|^3}
\label{eqn:g(r)}
\end{equation}
where $\bar{\rho}$ is the mean matter density and $\delta_m({\bf r})$
= $(\rho_m({\bf r})-\bar{\rho})/\bar{\rho}$ is the density contrast of
the mass perturbations.  In linear theory, the peculiar velocity
field, ${\bf v}({\bf r})$, and the peculiar acceleration vector, ${\bf
g}({\bf r})$, should be parallel with a constant of proportionality.

For ${\bf v}({\bf r})$, we use a value derived from the CMB dipole.
Using the first year of data from WMAP, \cite{Bennett} and the
revised values of the motion of the Sun relative to the LG a velocity
derived by \cite{Courteau}, we find a LG velocity
relative to the CMB of $v_{LG}=627\pm22$ \kms, towards
($l_{LG}=273^\circ\pm3^\circ$,$b_{LG}=29^\circ\pm3^\circ$).  The LG
dipole, ${\bf d}({\bf r})$ (in \kms), used as a proxy for ${\bf g}({\bf r})$, is
estimated from a galaxy survey and can be written as a summation of
all the galaxies in that survey (Equation 13 in \cite{erdogdu1}):
\begin{equation}
{\bf d}({\bf r})= \frac{H_0\Omega_{\rm m}^\gamma}{\rho_L b_L} \sum\limits_{i}^{N}\, w_{L_i}{\rm S}_{i} \hat{{\bf r}}_i\,,
\label{eqn:vlum}
\end{equation}
where $\rho_L$ is the luminosity density, $b_L$ is the luminosity bias factor introduced to account for
the dark matter haloes not fully represented by the galaxies
(hereafter we assume $b_L=1$). The index $\gamma$ appears in the
growth function $f = d \ln \delta /d \ln a \approx \Omega_{\rm m}^{\gamma}$
\cite{peebles} and is an important diagnostic for distinguishing
between $\Lambda$CDM and alternative models of gravity.  It has little
dependence on the cosmological constant \cite{lahav2}, and a
slight dependence on the dark energy equation of state parameters $w$
\cite{wang}. Recent refined calculations predict $\gamma
= 0.55$ for the $\Lambda$CDM concordance model, and $\gamma = 0.69$
\cite{linder} for a particular modified gravity model, DGP
braneworld gravity \cite{dvali}.

In equation~\ref{eqn:vlum}, $S_i$ is the flux of galaxy $i$ and $w_{L_i}$ is the luminosity weighting function written as:
\begin{equation}
w_{L_i} = \frac{1}{\int^{\infty}_{L_{\rm lim,i}}L\Phi (L)\, dL}\,,
\end{equation}
where $\Phi (L)$ is the luminosity function and $L_{\rm lim,i}$ is the minimum luminosity at a distance $r_i$
(see \cite{erdogdu1} for a full derivation). The Poisson shot noise of ${\bf d}$ is estimated as
\begin{equation}
\sigma_{sn}^2= \left(H_0\Omega_{\rm m}^{\gamma}\right)^2 \sum\limits_{i}^{N}\, \left(w_{L_i} S_i\hat{{\bf
r}}_i\right)^2\;.
\label{eqn:sn}
\end{equation}

For a whole sky survey, the non-zero elements of the covariance matrix are given as follows:
\be
C_{\rm dd}=\frac{1}{3}\lb \bd \cdot \bd \rb = \frac{H_0^2 \omegam^{2\gamma}}{3  (2\pi)^3} \int
\widehat{W}_{\bd}^2(k) P_{gg}(k) dk^3 \,
\label{eq:g^2}
\ee
and
\be
C_{\rm vv}=\frac{1}{3}\lb \bv \cdot \bv \rb = \frac{H_0^2 \omegam^{2\gamma}}{3  (2\pi)^3} \int
\widehat{W}_{\bv}^2(k) P_{mm}(k) dk^3 \,
\label{eq:gv}
\ee
and
\be
C_{\rm dv}=\frac{1}{3}\lb \bd \cdot \bv \rb = \frac{H_0^2 \omegam^{2\gamma}}{3  (2\pi)^3} \int
\widehat{W}_{\bv}(k)\widehat{W}_{\bd}(k) P_{gm}(k) dk^3 \, ,
\label{eq:v^2}
\ee
where $ P_{gg}$ and  $P_{mm}$ denote the galaxy and the dark matter power spectrum, respectively and  $P_{gm}$ denotes the galaxy-dark matter cross spectrum.

The expected misalignment angle between $\bv$ and $\bd$ is 
(JVW and LKH):
\begin{align}
p(\theta|\bv) 
\! & = \! \sin(\theta)\, \exp\left(-1/(2\theta_*^2)\right) 
\left\{\frac{y}{\sqrt{2\pi}} \right. \nonumber \\
\!\!\! & \hphantom{=} \!\!\! + \left(\frac{1+y^2}{2}\right) 
\exp\left(y^2/2\right) \left[1 + {\rm erf}(y/\sqrt{2})\right] 
\biggr\} \,,
\label{eq:te|v}
\end{align}
where $\theta$ is the angle between the two vectors, $\theta_*\equiv[\sqrt{1-c^2}/c]/(\bv/C_{\rm vv}^{1/2})$ (in radians)
and $c\equiv[C_{\rm dv}^{2}/(C_{\rm vv}C_{\rm dd})]^{1/2}$ and
$y\equiv\cos(\theta)/\theta_*$.  In the case where $\theta_*$ is
small, the probability distribution $p(\cos(\theta)|\bv)$ tends to
bi-variate Gaussian and $\theta_*$ measures the expected scatter. 

\subsection{Window Functions}

We assume the LG is a sphere with radius $r_{\rm LG}=1.6 \hmpc$ and 
the LG velocity is produced by the matter distribution out to horizon.
In Fourier space, 
the window function of the velocity vector ${\bv}$, determined from the CMB dipole, varies as 1/k:
\be
\widehat{W}_{\bv}(k) = \frac{j_0(k r_{\scriptscriptstyle \rm LG})}{k} \,.
\label{eq:W_v(k)}
\ee
For the flux-weighted dipole $\widehat{W}_{\bd}(k)$ can be written as:
\begin{eqnarray}
\widehat{W}_{\bd}(k)&=&4 \pi \! \int_{\rm r_{\rm LG}}^{\rm r_{\rm max}} r^2 dr W_{\bd}(r) j_1(kr) \nonumber \\
&=& \frac{j_0(k r_{\scriptscriptstyle \rm LG})}{k}- \frac{j_0(k r_{\scriptscriptstyle \rm max})}{k}\,,
\label{eq:W_d(k)}
\end{eqnarray}
where $W_{\bd}(r)=\frac{1}{4\pi r^2}$ and $r_{\rm max}$ is the maximum radius of the survey which we adopt as $r_{\rm max}= 130 \hmpc$. We want to compare our findings with the observed 
2MRS values and beyond this distance the shot-noise effects in the 2MRS becomes too large.

\subsection{Calculating the Power Spectra}
The {\it halo model} for non-linear
clustering gained popularity during the past few years  \cite{Ma,Peacock,Seljak}. This model assumes that most matter in the
Universe is structured as virialised halos and correlation function on
large scales depends on the clustering of these halos whereas 
non-linear clustering
of objects are related to the internal structure of the individual
halos. This implies that the non-linear power spectrum $P_{nl}(k)$ 
can be divided into a quasi-linear, {\it inter halo} term $P_{2h}(k)$ and a 
non-linear {\it intra halo} term $P_{1h}(k)$:
\be
P_{nl}(k)=P_{2h}(k)+P_{1h}(k).
\label{eqn.halo_pk}
\ee 

\subsection{The Dark Matter Power Spectrum}
Within the halo model framework, the dark matter power spectrum can be written as (e.g. \cite{Cooray}),
\begin{equation}
  \label{eqn.1halo.dm}
  P^{\rm{(1h)}}_{\rm{dm}}(k) = \int \rm{d}M~n(M)\left(\frac{M}{\bar{\rho}}\right)^2 |\hat{u}(k|M)|^2,
\end{equation}
and
\begin{equation}
\label{eqn.2halo.dm}
  P^{\rm{(2h)}}_{\rm{dm}}(k) =
  P^{\rm{lin}}_{\rm{dm}}(k)\left[\int\rm{d}M~n(M) b(M) \left(\frac{M}{\bar{\rho}}\right) \hat{u}(k|M)\right]^2,
\end{equation}
where the integrals are over the halo mass, $M$ and $\bar{\rho}$ is the background density of the universe. We use transfer functions of \cite{Eisenstein} to calculate 
$P^{\rm{lin}}_{\rm{dm}}(k)$. In the expressions above,
$n(M)$ is the halo mass function describes the number density of haloes of mass $M$, b(M) is the halo bias factor and $\hat{u}(k|M)$ is the Fourier
transform of the halo profile, $\rho(r)$.

For n(M) and b(M), we assume the forms proposed by \cite{Sheth}:
\begin{equation}
  \label{eqn.mf}
n(M)\rm{d}M = \frac{\bar{\rho}}{M}f(\nu)\rm{d}\nu,
\end{equation}
\begin{equation}
  \nu f(\nu) = A_{*} \left(1 + (q \nu)^{-p}\right)\left({q\nu\over 2\pi}\right)^{1/2}\rm{e}^{-q\nu/2},
\end{equation}
where $q=0.707$, $p=0.3$, and the normalisation $A_{*}\approx0.3222$. The mass variable is defined as $\nu\equiv\left(\delta_{\rm{sc}}(z)/\sigma(M)\right)^2$,
where $\delta_{\rm{sc}}(z)$ is the linear-theory prediction for the
present day overdensity of a region undergoing spherical collapse
at redshift $z$ ($\delta_{\rm{sc}}(0) \approx 1.68$) and $\sigma(M)$ is the r.m.s. variance of the present
day linear power spectrum in a spherical top-hat which contains an average mass
$M$. 

The degree of bias is also a function of the halo mass:
\begin{equation}
  \label{eqn.halo_bias}
  b(M) = 1 + \frac{q\nu-1}{\delta_{sc}(z)} + \frac{2p/\delta_{sc}(z)}{1 + (q\nu)^p}.
\end{equation}
We also use the \cite{nfw} dark matter halo density
profile.
\begin{equation}
\rho(r) = \frac{\rho_s}{(r/r_s)(1 + r/r_s)^2} \hspace{5mm} (r < r_{\rm
  vir})
\end{equation}
where $r_s$ is the characteristic scale radius and $\rho_s$ provides
the normalization.  The profile is truncated at the virial radius
$r_{\rm vir}$, which is obtained from the halo mass.
We parametrize the profile in terms of the {\it concentration
 parameter} $c = r_{\rm vir}/r_s$.  
The normalization for the mass
$M$ is
\begin{equation}
M = \int_0^{r_{\rm vir}} \rho(r) 4\pi r^2 dr = 4 \pi \rho_s r_s^3
\left[ \ln{(1+c)} - \frac{c}{1+c} \right]
\end{equation}
We assume that the concentration parameter $c$ depends on halo mass
$M$ and redshift in a manner calibrated by numerical simulations
(\cite{bullock, zehavi}):
\begin{equation}
c(M,z=0) = 11\left( \frac{M}{M_0} \right)^{-0.13},
\end{equation}
where $M_0$ is obtained from equation by setting the variance of the linear power spectrum $\sigma(M_0,0) = \delta_{\rm sc}=1.68$.  For our adopted cosmological model
we obtain $M_0 = 10^{12.13} \, h^{-1} M_\odot$.  

\subsection{The Galaxy Power Spectrum}
The components of the galaxy power spectrum are similar to the dark matter power spectrum given above 
\begin{equation}
  \label{eqn.1halo.gal}
  P^{\rm{(1h)}}_{\rm{gal}}(k) = \int \rm{d}M~n(M)\frac{\left< N(N-1)|M\right>}{\bar{n}_{\rm{gal}}^2} |\hat{u}(k|M)|^2,
\end{equation}
and
\begin{equation}
\label{eqn.2halo.gal}
  P^{\rm{(2h)}}_{\rm{gal}}(k) =
  P^{\rm{lin}}_{\rm{dm}}(k)\left[\int\rm{d}M~n(M) b(M) \frac{\left< N|M\right>}{\bar{n}_{\rm{gal}}}\hat{u}(k|M)\right]^2,
\end{equation}
Here, $\left<N|M\right>$ and $\left<N(N-1)|M\right>$ are the first and second factorial moments of the
halo occupation distribution, $P(N|M)$, respectively,
$\bar{n}_{\rm{gal}}$ is the average number density of galaxies.
The halo occupation distribution (HOD) is a parametrized description
of how galaxies populate dark matter haloes as a function of the halo
mass $M$. We
adopt simple models motivated by results from simulations and
semi-analytic calculations (e.g. \cite{Kravtsov})
\begin{equation}
  \label{eqn.hod}
  \langle N | M \rangle =\left\{ \begin{array}{ll}
 \left(M/M_0\right)^\beta & (M \ge \rm{M_{cut}}) \\
 0 & (M < \rm{M_{cut}}) 
  \end{array} \right.
\end{equation}
where $M_0$ and $\beta$ are free parameters and $\rm{M_{cut}}$ is determined by matching $\bar{n}_{\rm{gal}}$ to 
the value derived from the 2MRS ($\bar{n}_{\rm{gal}}=7.7\times10^{-4} h^3{\rm Mpc}^{-3}$).
The second moment of $P(N|M)$ is:
\begin{equation}
  \label{eqn.hod_sec}
  \langle N(N-1)|M \rangle = \left\{ \begin{array}{ll}
    \langle N | M \rangle^2 & (M \ge 10^{13}h^{-1} \msun ) \\
  (\alpha \langle N | M \rangle)^2 & \mbox{otherwise}\,, 
  \end{array} \right.
\end{equation}
where $\alpha=log\sqrt{M/h^{-1}10^{11} \msun}$.

For our halo model parameters, we use $M_0=10^{14.16} {h}^{-1} \msun$ and $\beta=0.85$, derived for 2MASS galaxies (see Table 1 of \cite{Lin}).
These values are quite typical for red galaxies and varying the parameters slightly 
does not affect the results.  

We note that on large scales the galaxy power spectrum takes the simple form:
\begin{equation}
  \label{eqn.simple_pk}
P_{\rm gal}(k) = b^2 P_{\rm dm}^{\rm lin}(k) + 1/{\bar n}_{\rm gal}
\end{equation}
The last term is the shot noise term, which we found to give the very r.m.s.
contribution to the shot-noise in the dipole as the empirical estimate (Equation~\ref{eqn:sn}).

\subsection{The Galaxy-Dark Matter Cross Power Spectrum}
Using the expressions above, the galaxy dark matter cross power spectrum components can be written as 
\begin{equation}
  \label{eqn.1halo.gm}
  P^{\rm{(1h)}}_{\rm{gm}}(k) = \int \rm{d}M~n(M)\frac{M}{\bar{\rho}}\frac{\left< N|M \right>}{\bar{n}_{\rm{gal}}} |\hat{u}(k|M)|^2,
\end{equation}
and
\ba
\label{eqn.2halo.gm}
  P^{\rm{(2h)}}_{\rm{gm}}(k) &=& P^{\rm{lin}}_{\rm{dm}}(k)\left[\int\rm{d}M~n(M) b(M) \frac{\left< N|M\right>}{\bar{n}_{\rm{gal}}}\hat{u}(k|M)\right] \nonumber \\
  &\times& \left[\int\rm{d}M n(M) b(M) \left(\frac{M}{\bar{\rho}}\right) \hat{u}(k|M)]\right].
\ea

\section{Convergence}
We now turn to the question of the convergence of the acceleration vector derived from a galaxy survey given the velocity of the CMB. We recalculate the conditional probability,$p(\bd|\bv)$ given in equation~\ref{eq:te|v} within a series of successively larger concentric spheres centred on the LG. In this case, the value of the covariance matrix element $C_{\rm vv}$ remains the same and we recalculate $C_{\rm dv}$ and  $C_{\rm dd}$ using a series of window functions: 
\be
\widehat{W}_{\bd}(k) = \frac{j_0(k r_{\scriptscriptstyle \rm LG})}{k}- \frac{j_0(k R)}{k}\,.
\ee
where $R$ is the outer limit of each shell. 

The expected dipole velocity given the CMB dipole is then written as:
\be
\bd(R)=\frac{C_{\rm dv}(R)}{C_{\rm vv}}\bv_{\rm CMB}
\label{eq:conv}
\ee
and the `1 sigma' scatter in the velocity is $|\bd(R)|\pm \sqrt{3\sigma^2}$ with 
$\sigma^2 = C_{\rm dd}(R) (1-\frac{C_{\rm dv}(R)^2}{C_{\rm dd}(R)C_{\rm vv}})$.

\section{Results}

\begin{figure}
\psfig{figure=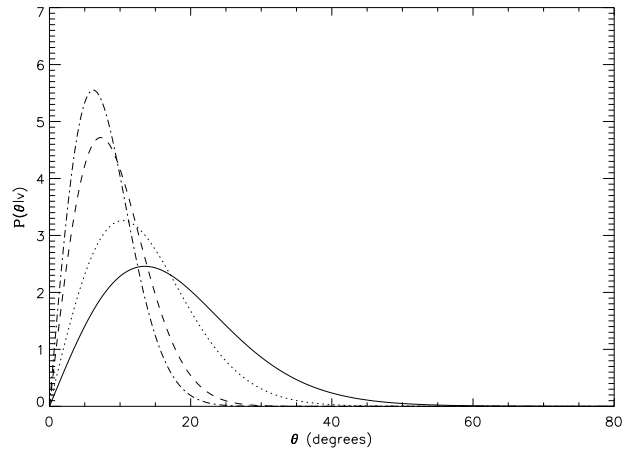,angle=0,width=0.5\textwidth, clip=} 
\caption{The probability distribution functions for misalignment 
angle $\Theta$ between the galaxy survey dipole and the CMB velocity vectors.
The dashed-dotted
line denotes the misalignment angle calculated using $\rm P(k)_{\rm mm}^{lin}$ only for SCDM model ((1) in Table 1), the dashed line is calculated using $\Lambda$CDM and $\rm P(k)_{\rm mm}^{nl}$ ((3) in Table 1). The solid line is calculated using Equations~\ref{eq:g^2},~\ref{eq:gv} and ~\ref{eq:v^2} ((4) in Table 1).  The dotted line is the same as the solid line but 
using $\rm P(k)_{\rm gg}= \rm P(k)_{\rm gg}^{2h}$ ((5) in Table 1).}
\label{fig:diprob}
\end{figure}

In Figure~\ref{fig:diprob}, we plot a family of $p(\theta|\bv)$ curves given by equation~\ref{eq:te|v}. 
The dashed-dotted line is calculated using the old SCDM model with  $\Omega_{\rm m}=1.0$,
$\sigma_8=0.6$, the Hubble constant $H_0=50 \kms {\rm Mpc}^{-1}$ and the spectral index $n=1$.
For all other power spectra, we use WMAP5
fiducial parameters $\Omega_{\rm m}=0.26$, $f_b=0.17$,
$\sigma_8=0.79$, $H_0=72 \kms {\rm Mpc}^{-1}$ and $n=1$  \cite{Dunkley}. 

We first consider the standard linear and non-linear $\Lambda$CDM dark matter power spectrum models with a linear
biasing model of $b=1$.  We obtain almost identical curves (dashed
line) with a scatter of $\theta_*=7^\circ$.  Both curves peak at
around $8^\circ$. Changing the radius of the LG to $r_{\rm LG}=5
\hmpc$ has also a very small effect on the results. The narrowest
scatter in the probability curve and the smallest expected
misalignment angle is obtained using the SCDM model
($\theta_*=6^\circ$, dashed-dotted line). This is expected since SCDM
has a smaller coherence length than the $\Lambda$CDM model.

\begin{figure}
\psfig{figure=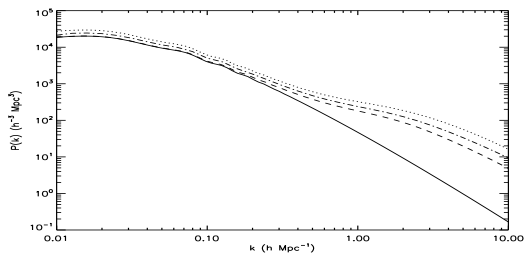,angle=0,height=35mm,width=0.4\textwidth, clip=} 
\caption{The linear dark matter ($P_{\rm lin}$, solid line), halo model dark matter ($P_{\rm dm}$, dashed line), galaxy-dark matter ($P_{\rm gm}$, dashed-dotted line) and galaxy-galaxy ($P_{\rm gal}$, dotted line) power spectra. The power spectra are normalized
to be equal on very large scales (very small k).}
\label{fig:pk}
\end{figure}

We then calculate the power spectra using the halo model for galaxies
as well as for dark matter. 
The power spectrum shown by the black solid line includes
both 1-halo and the 2-halo terms. 
The scatter for this model is the largest
$\theta_*=14^\circ$ and the probability curve peaks at the largest
misalignment angle of the plotted curves $(13.5^\circ)$ . In the
galaxy halo model, the galaxies are more clustered than the dark
matter (see Figure~\ref{fig:pk}) 
and thus the covariance matrix elements are much larger than
the ones we obtain with the linear bias models.
The halo model is used to describe all haloes within a survey. 
Thus the 2-halo term represents the clustering of all haloes on large scales and the 1-halo term is related to the internal clustering of each halo in the survey.
We also want to model the dipole induced by clusters of galaxies
rather than galaxies themselves. For this, we plot the probability
function calculated using 2-halo term only ($P_{gg}(k)=P_{2h}(k)$, the
dotted line) which peaks around $10^\circ$ and gives
$\theta_*=11^\circ$.  When we add the 1-halo term (solid line), we are
adding the galaxy component to our model. This increases the
shot-noise which causes the CMB and the galaxy survey dipoles to
decorrelate as confirmed by our value for $\theta_*$. Changing the radius of the LG to $r_{\rm LG}=5
\hmpc$ mitigates the non-linear effects, reducing the scatter, $\theta_*=10^\circ$.

We also calculate the r.m.s. velocity of our LG models, $<{\bf v}_{\rm
LG}^2>^{1/2}=(3C_{vv})^{1/2}$. For our $\Lambda$CDM cosmology with $\sigma_8=0.79$, we
obtain $~412$ and $~430$ \kms for the linear and non-linear power
spectra, respectively.  We note the observed LG velocity amplitude, 621 \kms, is larger than these r.m.s. values. 

We summarize the results presented in
Figure~\ref{fig:diprob} in Table 1.  We present the probability
obtaining the actual observed angle ($21^\circ$) and the integrated
probability for $\theta$ to lie below the observed value which
provides a frequentist test. The two columns indicate that the
observed dipole is highly unlikely in both the linear and non-linear
models which do not include a non-linear bias model for
galaxies. Neither are we able to approximate the observed value
without including the 1-halo term to the galaxy power spectrum. But,
using the halo model for both galaxies and the dark matter, results in
a mean close to that of $21^\circ$.

\begin{table*}
\caption[]{Dipoles constrained by the velocity of the local group, the characteristic misalignment angle $\theta_*$ is given in degrees. The observed misalignment angle $\theta_{obs}=21^\circ$ is calculated from the 2MRS survey (Erdo{\u g}du \etal 2006b). The assumed cosmological parameters ($\sigma_8$,$h_0$,$\Omega_{\rm m}$,$f_b$) are (0.79,0.72,0.26,0.17) for $\Lambda$CDM and (0.6,0.5,1.0,0.044) for SCDM.}
\begin{center}
\begin{tabular}{@{}llll}
\hline \\ Model & $\theta_* (deg)$ & $P(\theta_{\rm obs})$/max($P(\theta|\bv))$ & Percent below $\theta_{\rm obs}$ \\\hline
(1) SCDM, $P^{lin}_{\rm dm}(k)$ & 6.3 & 0.02 & 99.6 \\ 
(2) $\Lambda$CDM, $P^{lin}_{\rm dm}(k)$  & 7.2 & 0.08 & 98.1\\
(3)  $\Lambda$CDM, $P^{nl}_{\rm dm}(k)$  & 7.2 & 0.08 & 98.1 \\
(4)  $\Lambda$CDM, Halo model for Galaxies& 13.5 & 0.78 & 67.2 \\ 
(5)  $\Lambda$CDM, $P_{\rm gg}(k)=P^{2h}_{\rm gg}(k)$ & 10.4 & 0.45 & 85.3 \\ 
\end{tabular}
\end{center}
\end{table*}
 
\begin{figure}
$\begin{array}{cc}
\psfig{figure=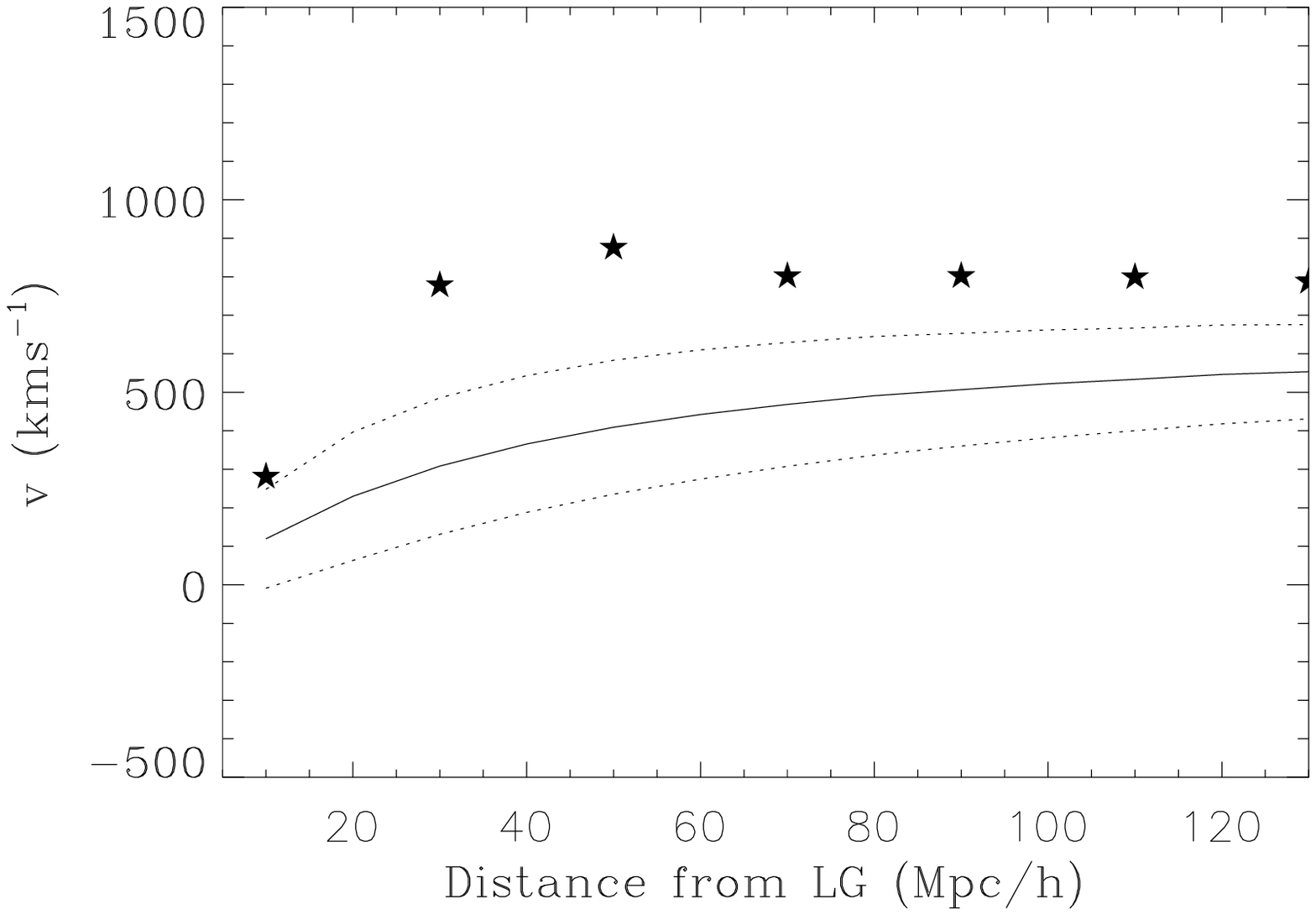,angle=0,width=0.25\textwidth,clip=} 
\psfig{figure=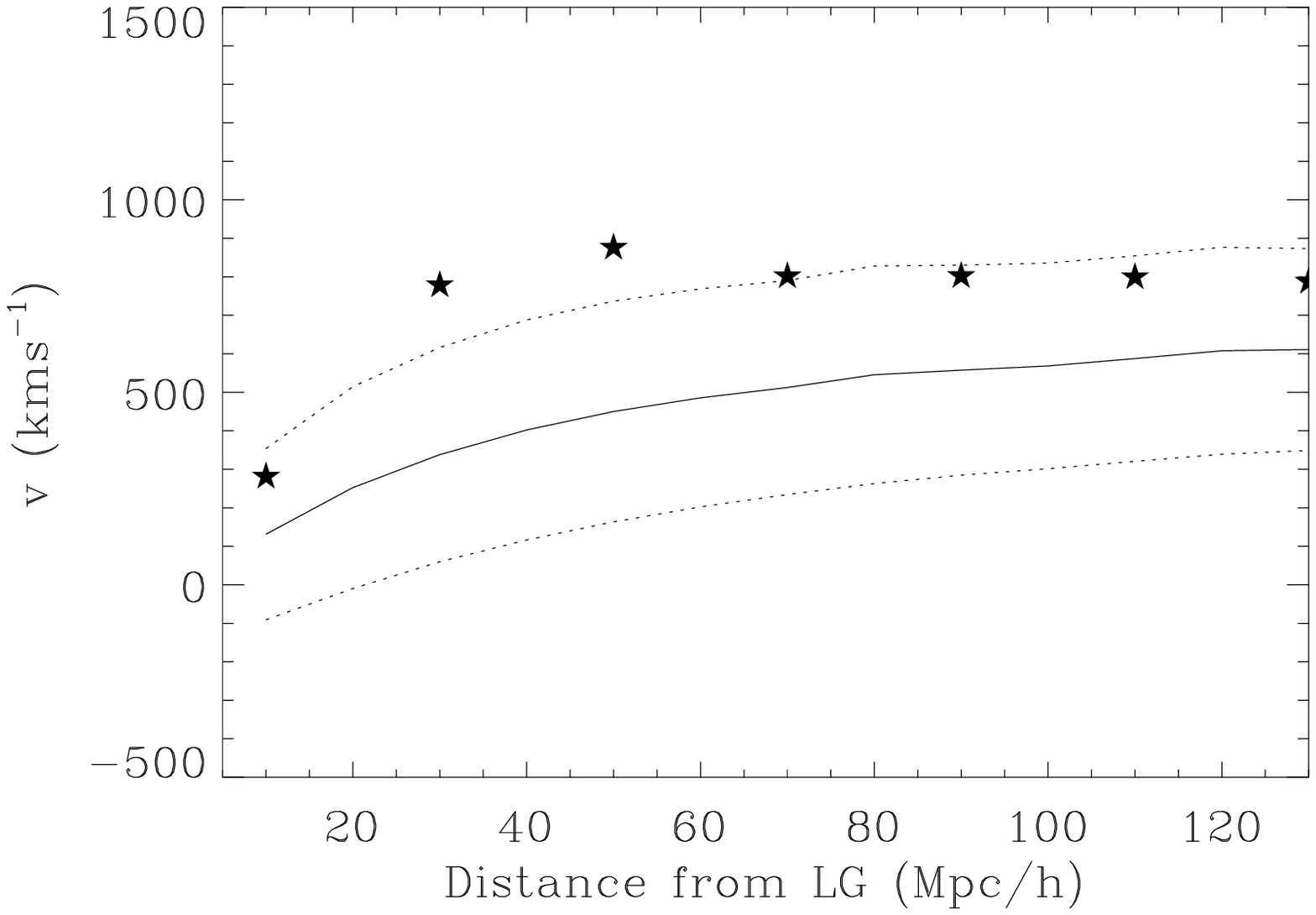,angle=0,width=0.25\textwidth,clip=} \\
\psfig{figure=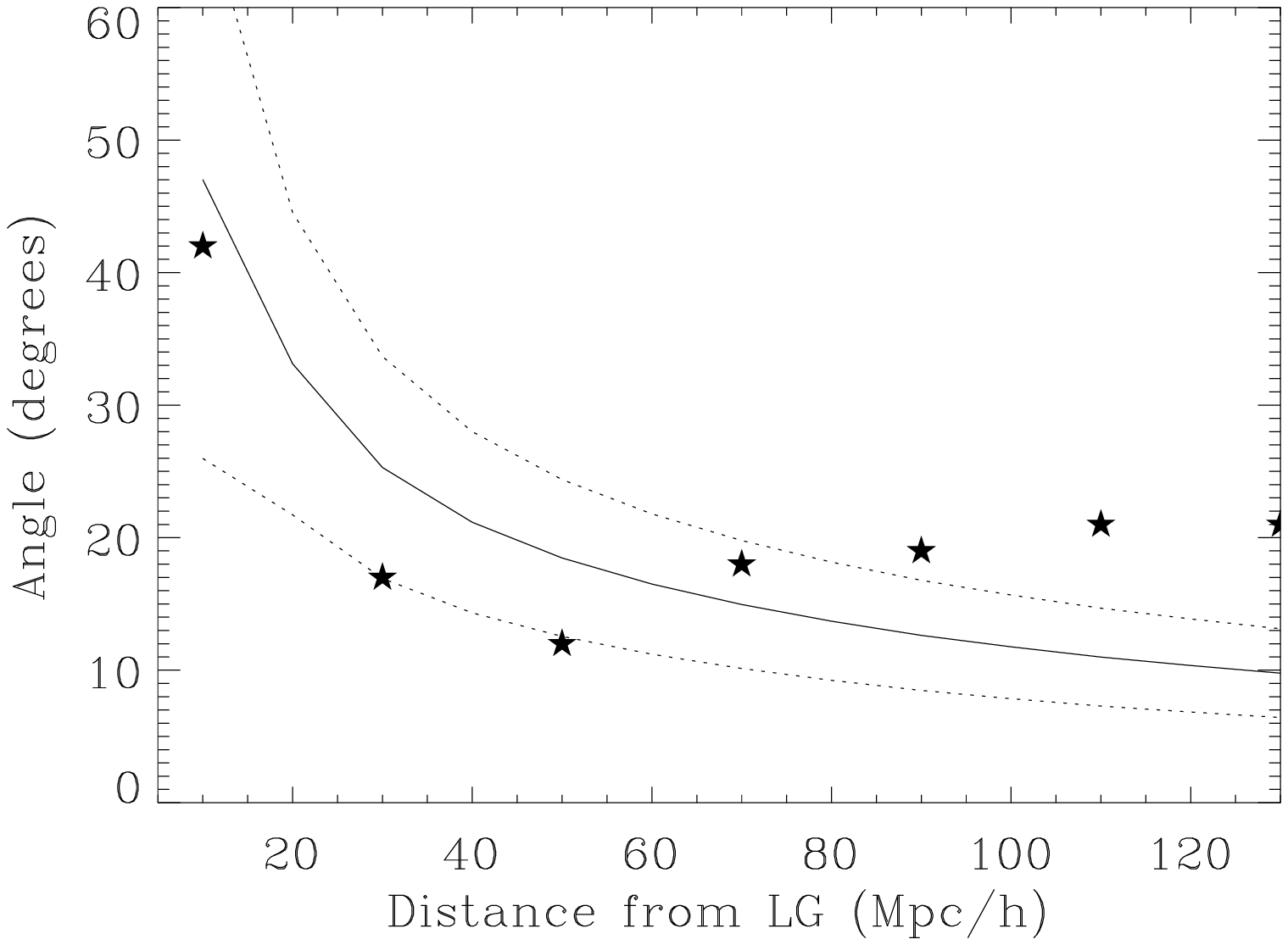,angle=0,width=0.25\textwidth,clip=} 
\psfig{figure=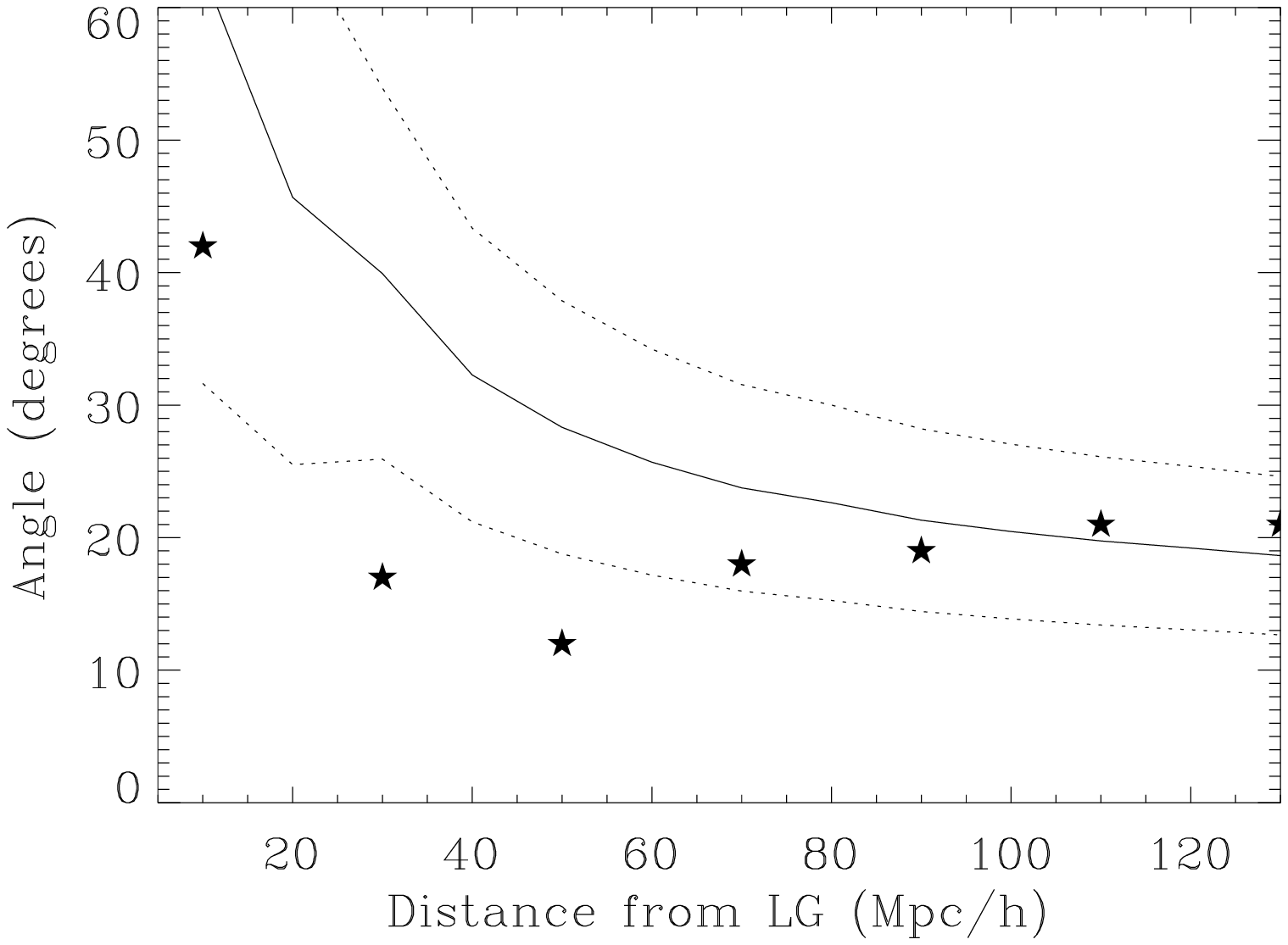,angle=0,width=0.25\textwidth,clip=} \\
\end{array}$
\caption{{\bf Top Panel}: The predicted LG velocity within a series of successively larger concentric spheres centred on the LG given by Equation~\ref{eq:conv} and the 1-sigma scatter. {\bf Bottom Panel:} The misalignment angle as a function of distance. {\bf Left}: The convergence is calculated using linear dark matter power spectrum only ((2) in Table 1). {\bf Right}: The convergence is calculated using non-linear galaxy power spectrum and galaxy-dark matter cross spectrum (Case (4) in Table 4).}
\label{fig:conv}
\end{figure}

In Figure~\ref{fig:conv}, we plot the expected dipole velocity given
the CMB dipole (Equation~\ref{eq:conv}, solid line) and its scatter 
(dashed line).  In the top panel, we use only
the linear power spectrum (the first model in Table 1) and the bottom
panel is the velocity convergence with the full halo model treatment
(the fourth model in Table 1). The plots in the left panel are the
predicted amplitudes of the acceleration on the LG as a function of
distance. The right plots show the convergence of the angle between
the galaxy survey dipole and the CMB dipole. We also plot the observed
values of the acceleration amplitude and the misalignment angle from
the 2MRS survey obtained using the flux-weighted selection function
(\cite{erdogdu1}) as an illustrative example. 
There are many determinations of the dipole misalignment angle from the 2MRS data ranging from 14$^\circ$ \cite{erdogdu1} to 50$^\circ$ \cite{lavaux} depending on the type of analysis performed. The near-infrared flux weighted dipoles are more robust that the number weighted schemes because they closely approximate a mass-weighted dipole, bypassing the effects of redshift distortions and require no preferred reference frame. Our analysis holds for either scheme or the choice of reconstruction technique since we account for the correct data vector for selection effects 
and compare like with like. 

The 2MRS acceleration grows rapidly out
to $\approx 50 \hmpc$ after that it flattens off. The misalignment
angle drops to $12^\circ$ at $\approx 50 \hmpc$ but then increases at
larger distances. This behaviour is not seen for either of the
models. This is not unexpected, however, since the increase in
amplitude at $\approx 50 \hmpc$ and than its consequent decrease at
$\approx 60 \hmpc$ is attributed to the tug-of-war between the Great
Attractor and the Pisces-Perseus Supercluster which are not accounted
in the statistical (r.m.s.) models. 

In theory, one can equate the velocity
inferred from the CMB measurements with the value derived from a
galaxy survey and obtain a value for $\omegam$.  In practice, however,
the galaxy surveys do not measure the true total velocity due their
finite depth and the predicted amplitude of
the reconstructed velocity depends the 
combination of matter density and the biasing parameter $b$ 
which is usually not known to great accuracy. However, one would expect
$\omegam$ and $b$ inferred from the LG velocity to have observationally viable values.
Figure~\ref{fig:conv} shows that the amplitudes for both of the models
are similar to observations, but the non-linear model has higher
amplitude due to higher small-scale power and is closer to the
observed values. In both models, more than half of the predicted LG velocity signal is generated within $\approx 30 \hmpc$.
The non-linear galaxy model decorrelates the angles and
incorporates shot-noise effects and closely resembles the observed
convergence.

\section{Discussion}
We have compared estimates for of the acceleration as derived from a 2MRS like survey and the motion of the LG in a $\Lambda$CDM universe both within and without the halo model framework.
We showed that a combination of linear theory with the halo model
predicts convergence and misalignment very much in agreement with the
observed 2MRS dipole.

We should note our analysis assumes linear perturbation theory but we are using a non-linear galaxy power spectrum,
so our treatment is incomplete. A more consistent way would be to use N-body simulations 
(e.g. \cite{DavisStrauss})
and semi-analytic models galaxy formation but this is beyond the scope of the work presented in this paper. 
Here, we tested if a non-linear power spectrum model with a more sophisticated biasing formula could predict the observed acceleration better than a linear power spectrum.
Indeed, our results indicate the flow in our location in the universe
is quite typical for a $\Lambda$CDM model.  A non-linear velocity-acceleration relation will increase the misalignment angle \cite{DavisNusser}. Recently, Lavaux \etal \cite{lavaux} used the 2MRS reconstructed peculiar velocity field to estimate a LG velocity misalignment of $\approx50^\circ\pm22^\circ$ (at 95\% confidence). 
They suggest the dipole directions should agree to within $25^\circ$ at 95\% confidence. From Figure 3 (right panel), the expected misalignment angle is $\approx29^\circ\pm7^\circ$ at 130 \hmpc. Our analysis should be viewed as a conservative estimate of the expected misalignment error.  Thus, even a dipole misalignment is as high as reported by \cite{lavaux} is reconcilable with a $\Lambda$CDM model given their estimated error.

We would like to emphasize that we have assumed that the accelerations are estimated from a whole sky survey. It is possible to incorporate the effect of ZoA (see Appendix of LKH) which would increase the scatter substantially. We note that we are able to reproduce the observed values very well without modeling missing structures in the masked regions as suggested by \cite{Loeb}. We further note that 
many dedicated searches of the ZoA did not detect such 
massive structures (see e.g. \cite{Ebeling1, Ebeling2, Kocevski2, Kocevski3, 
Roman, Wakamatsu,Hasegawa, Meyer,
Henning, KraanKorteweg, KraanKorteweg2}). Our analysis should be understood in the r.m.s. sense. It does not pick out directions. As noted by others (e.g. \cite{Loeb}), most of the LG misalignment lies along the direction of the galactic centre, the region of most obscuration, and this is another route for exploring the misalignment.

To summarize, we presented a fresh analysis of the fundamental
problem of the origin of motion on the Local Group. We conclude that the newly observed 2MRS dipole is in accord with what we would expect in a $\Lambda$CDM universe within the context of the halo model.

\section {Acknowledgements} 
We thank Michal Chodorowski, Gert Hutsi and Ren\'ee Kraan-Korteweg for helpful discussions. 
PE would like to thank UCL for its hospitality. OL acknowledges the support of a Wolfson Royal Society Research Merit Award.


\begin{thebibliography}{99}

\bibitem{Yahil}Yahil A., Walker D. \& Rowan-Robinson M., 1986, ApJ, 301, L1
\bibitem{Meiksin}Meiksin A. \& Davis M., 1986, AJ, 91, 191
\bibitem{Harmon}Harmon R.T., Lahav O. \& Meurs E.J.A., 1987, MNRAS, 228, 5
\bibitem{Villumsen}Villumsen J.V.  \& Strauss M.A., 1987, ApJ, 322, 37
\bibitem{lahav1}Lahav O., Rowan-Robinson M. \& Lynden-Bell, 1988, MNRAS,
234, 677
\bibitem{lynden}Lynden-Bell D., Lahav O. \& Burstein D., 1989, MNRAS,
241,325
\bibitem{Strauss} Strauss M. A., Yahil A., Davis M., Huchra J. P., Fisher K., ApJ, 1992, 397, 395
\bibitem{Webster}Webster M., Lahav O. \& Fisher K., 1997, MNRAS, 287,425
\bibitem{LKH}Lahav O., Kaiser N., Hoffman Y., 1990, ApJ, 352, 448 
\bibitem{JVW}Juszkiewicz R., Vittorio N., Wyse R., 1990, ApJ, 352, 408
\bibitem{erdogdu1}Erdo{\u g}du P., et al., 2006a, MNRAS, 368, 1515 
\bibitem{Kaiser}Kaiser N., 1987, MNRAS, 227, 1
\bibitem{erdogdu2}Erdo{\u g}du P., et al., 2006b, MNRAS, 373, 45
\bibitem{lavaux}Lavaux G., Tully R. B., Mohayaee R., Colombi, S., 2008, arXiv:astro-ph/0810.3658v2
\bibitem{crook} Crook A., Silvestri A., Zukin P., 2009, arXiv:astro-ph/0906.2411
\bibitem{Loeb}Loeb A., Narayan R., 2008, MNRAS, 386, 2221
\bibitem{Chodorowski}Chodorowski M. J., Coiffard J., Bilicki M., Colombi S., Ciecielag P., 2008, astro-ph/0706.0619
\bibitem{Kashlinsky}Kashlinsky A., Atrio-Barandela F., Kocevski D., Ebeling H., 2008, ApJ, 686, 49
\bibitem{Watkins}Watkins R., Feldman H. A.,Hudson M. J., 2009, MNRAS, 392, 743
\bibitem{Bennett}Bennett C.L., \etal, 2003, ApJS, 148, 1
\bibitem{Courteau}Courteau S. \& Van Den Bergh S., 1999, AJ, 118, 337
\bibitem{peebles}Peebles P. J. E., The Large-Scale Structure of the Universe, 
Princeton: Princeton Univ. Press, Princeton, N.J., USA 
\bibitem{lahav2} Lahav O., Lilje P. B., Primack J. R., Rees M. J., 1991,
MNRAS, 251, 128
\bibitem{wang}Wang L., Steinhardt P. J., 1998, ApJ, 508, 483
\bibitem{linder} Linder E. V., Cahn R. N., 2007, APh, 28, 481
\bibitem{dvali}Dvali G., Gabadadze G., Porrati M., 2000, PhLB, 485, 208
\bibitem{Ma}Ma C., Fry J. N., 2000, ApJ, 543, 503
\bibitem{Peacock}Peacock J.A., Smith R.E., 2000, MNRAS, 318, 1144
\bibitem{Seljak}Seljak U., 2000, MNRAS, 318, 203
\bibitem{Cooray}Cooray A., Sheth R.~K., 2002, Phys. Rep., 371, 1
\bibitem{Eisenstein}Eisenstein D. J., Hu W., 1998, ApJ, 496, 605
\bibitem[\protect\citeauthoryear{Sheth \& Tormen}{1999}]{Sheth}
    Sheth R.~K., Tormen G., 1999, MNRAS, 308, 119
\bibitem{nfw}Navarro J. F., Frenk C. S., White S. D. M., 1997, ApJ, 490, 493
\bibitem{bullock}Bullock J. S., Kolatt T. S., Sigad Y., Somerville R. S., Kravtsov A. V., Klypin
A. A., Primack J. R., Dekel A., 2001, MNRAS, 321, 559
\bibitem{zehavi} Zehavi I. et al., 2004, ApJ, 608, 16
\bibitem[\protect\citeauthoryear{Kravtsov et~al.}{2004}]{Kravtsov}
    Kravtsov A.~V., Berlind A.~A., Wechsler R.~H., Klypin A.~A.,
    Gottl\"ober S., Allgood B., Primack J.~R., 2004, ApJ, 609, 35
\bibitem{Lin}Lin Y., Mohr J.J., Stanford S.A., 2004, ApJ, 610, 745
\bibitem{Dunkley}Dunkley J. \etal 2009, ApJ Supp., 180, 306  
\bibitem{DavisStrauss}Davis M., Strauss M.A., Yahil A., 1990, ApJ, 372, 394
\bibitem{DavisNusser}Davis M., Nusser A., Willick J., 1996, ApJ, 473, 22
\bibitem{Ebeling1} Ebeling, H., Mullis, C. R., Tully, R. B. 2002, ApJ, 580,
774
\bibitem{Ebeling2}Ebeling H., Kocevski D., Tully R. B.,  Mullis C. R.
2005, Nearby Large-Scale Structures and the ZoA, 329, 83
\bibitem{Kocevski2} Kocevski, D. D., Ebeling, H., Mullis, C. R.,
Tully, R. B. 2005, ArXiv Astrophysics e-prints,
arXiv:astro-ph/0512321
\bibitem{Kocevski3}Kocevski D. D., Ebeling H., 2006, ApJ, 645, 1043
\bibitem{Roman} Roman, A. T., Takeuchi, T. T., Nakanishi, K., Saito, M.
1998, PASJ, 50, 47
\bibitem{Wakamatsu}Wakamatsu K., Malkan, M. A., Nishida M. T., Parker
Q. A., Saunders W., Watson F. G. 2005, Nearby Large-
Scale Structures and the ZoA, 329, 189
\bibitem{Hasegawa}Hasegawa T., et al. 2000, MNRAS, 316, 326
\bibitem{Meyer}Meyer M. J., et al. 2004, MNRAS, 350, 1195
\bibitem{Henning}Henning P. A., Kraan-Korteweg R. C., Stavely-Smith
L. 2005, Nearby Large-Scale Structures and the ZoA, 329, 199
\bibitem{KraanKorteweg}Kraan-Korteweg R. C., Lahav 2000 A\&A Rev, 10, 211
\bibitem{KraanKorteweg2}Kraan-Korteweg R. C., Shafi N., Koribalski B., Staveley-
Smith L., Buckland P., Henning, P. A., Fairall, A. P.
2007, arXiv:0710.1795
\end{thebibliography}
\end{document}